# Report on Female Participation in Informatics degrees in Europe

Report for the Eugain Virtual Mobility Grant

Grantee:
>   Andrea D'Angelo, University of L'Aquila

Participants:
>   Tiziana Catarci, Sapienza University of Rome
>   Antinisca Di Marco, University of L'Aquila
>   Monica Landoni, Università della Svizzera italiana, USI
>   Enrico Nardelli, Tor Vergata University of Rome
>   Giovanni Stilo, University of L'Aquila

## Disclaimer

The binary distinction between female and male students is based on the limitations of the available data. We recognize the importance of inclusion, particularly regarding gender diversity and, more broadly, intersectionality. If more comprehensive data were available, it would have allowed for richer and more meaningful analyses.

## 1. Summary

This study aims to enrich and leverage data from the Informatics Europe's Higher Education (IEHE) data portal [4] to extract and analyze trends in female participation in Informatics across Europe. The research examines the proportion of female students, first-year enrollments, and degrees awarded to women in the field. The issue of low female participation in Informatics has long been recognized as a persistent challenge and remains a critical area of scholarly inquiry [2,3]. Furthermore, existing literature indicates that socio-economic factors can unpredictably influence female participation, thereby complicating efforts to address the gender gap [1].

The analysis focuses on participation data from research universities at various academic levels, including Bachelor's, Master's, and PhD programs, and seeks to uncover potential correlations between female participation and geographical or economic zones. To this end, the dataset was first enriched by integrating additional information, such as each country's GDP and relevant geographical data, sourced from various online repositories, as detailed in Section 2.1. Subsequently, the data was cleaned to ensure consistency and eliminate incomplete time series. The final set of complete time series selected for further analysis is presented in Section 2.2.

We then used the data collected from the internet to assign countries to different clusters. Specifically, we individually employed the Economic Zone (Section 3.1), the Geographical Area (Section 3.2), and, lastly, the GDP quartile (Section 3.3) to cluster countries and compare their temporal trends between countries in the same class and against other classes. We analyze the results for each classification and derive conclusions based on the available data.

The study concludes with an analysis and interpretation of the data, using visual representations based on clustering variables, specifically economic zones, geographical areas, and GDP, to offer insights into the observed trends.

# 2. Data Collection and Clustering

## 2.1 Data Collection

The initial phase of this project focused on enriching the dataset from the Informatics Europe's Higher Education (IEHE) data portal by incorporating supplementary information from various publicly available and reputable databases. These additional data provided a richer context for analysis and facilitated the classification and clustering of countries.

Firstly, GDP per capita was sourced from the World Bank, a well-established international financial institution that provides comprehensive economic data for countries worldwide, offering insights into each country's economic performance by measuring the average income per person, adjusted for purchasing power parity.

Additionally, the classification of countries into different economic zones was obtained from Informatics Europe, the association of European university departments, and the industrial research lab in informatics, whose goal is advancing research and education in informatics across Europe. This classification provides a deeper understanding of how countries are grouped based on their economic frameworks and regional affiliations, which can influence higher education and technology sectors.

Area and population data were gathered from Wikipedia, which provides regularly updated demographic and geographic information. These figures include the total land area of each country and its corresponding population size, giving a foundational understanding of the physical and demographic scale of the nations in question.

Latitude and longitude coordinates, specifically for the centroid of each country, were obtained from GeoPandas, a geospatial data analysis tool widely used in academic research. These coordinates enable the precise geographical mapping of countries, allowing for spatial analyses and comparisons.

Lastly, geographical area data were sourced from the United Nations Statistics Division (UNSD), which provides a standard classification system used to define the regions and subregions of the world. This source is critical for ensuring the analysis adheres to globally recognized geographical boundaries and regional classifications.

The details of the acquired data are reported in Table 1.

| Collected Information | Source |
|---|---|
| GDP Per Capita | https://data.worldbank.org/indicator/NY.GDP.PCAP.PP.CD?end=2023&start=2000 |
| Economic Zone | https://www.informatics-europe.org/join-us/why-how-join.html |
| Area and Population | Wikipedia.org (for instance, https://it.wikipedia.org/wiki/Italy) |
| Latitude and Longitude of the Centroid of each Country | https://geopandas.org/en/stable/ |
| Geographical Area | https://unstats.un.org/unsd/methodology/m49/ |

**Table 1:** Additional Sources of Information

## 2.2 Data Cleaning and Convention

The initial dataset included records with incomplete data. For this reason, we need to analyze the data to keep only valid records. To do so, we retain only a complete time series with data available every year from 2010 to 2022. Specifically, any time series containing at least one entry marked as "tbp" (to be published) or "nan" (not available) was removed. It is important to note that this study focuses exclusively on data from research universities, which ensures a fairer comparison, as some European countries lack applied sciences universities. The comprehensive list of time series utilized in the analysis is provided in Table 2. In particular, the column 'Students' refers to the total number of enrolled students across each academic program (Bachelor, Master, PhD) in a certain year, column 'First Year Students' denotes those in their initial year of study in a certain year, and column 'Awarded Degrees' represents the total number of degrees conferred within each program in the reference year.

|  | **Students** | | | **First Year Students** | | | **Awarded Degrees** | | |
|---|---|---|---|---|---|---|---|---|---|
|  | **Bach** | **Mast** | **PhD** | **Bach** | **Mast** | **PhD** | **Bach** | **Mast** | **PhD** |
| **Austria** | X | X | X | X |  |  | X | X | X |
| **Belgium** | X | X |  | X |  |  |  |  |  |
| **Bulgaria** |  |  |  |  |  |  | X | X | X |
| **Czechia** |  |  |  |  |  |  |  |  |  |
| **Denmark** | X | X |  | X |  |  | X | X |  |
| **Estonia** | X | X | X | X |  |  | X | X | X |
| **Finland** | X | X | X | X |  |  | X | X | X |
| **France** |  |  |  |  |  |  |  |  |  |
| **Germany** | X | X | X | X |  |  | X | X | X |
| **Greece** |  |  |  |  |  |  |  |  |  |
| **Ireland** | X | X | X | X |  |  | X | X | X |
| **Italy** |  | X |  |  |  |  | X | X |  |
| **Latvia** | X | X | X |  |  |  | X | X | X |
| **Lithuania** | | | | | | | | | |
| **Netherlands** | X | X | | X | | | X | X | |
| **Norway** | X | X | | X | | | X | X | |
| **Poland** | X | X | | | | | | | |
| **Portugal** | X | X | X | X | | | X | X | X |
| **Romania** | X | | | | | | X | | |
| **Spain** | X | X | | | | | X | X | X |
| **Switzerland** | X | X | X | X | | | X | X | X |
| **Turkey** | | | | | | | X | X | X |
| **UK** | X | X | | X | | | X | | X |

**Table 2:** Complete Time Series for analysis.

Table 2 indicates that data for "First Year Students" is only available for bachelor's programs, leaving a gap in registration information for master's and PhD programs. Additionally, some countries (namely Czechia, France, Greece, and Lithuania) do not provide any complete time series and will therefore be excluded from further analysis. On the other hand, data on Awarded Degrees is generally more comprehensive than enrollment data.

# 3. Temporal Trends

After cleaning the dataset, we analyzed the temporal trends for each available and complete time series (see Table 2). Each time series is characterized by several attributes, including the country, economic zone, geographic area, statistical type (stat_type), and grade level. The country denotes the reference nation, while the economic zone is classified into three categories (0, 1, or 2), and the geographic area is categorized as North, South, West, or East (refer to Table 1 for data sources).

The *"stat_type"* attribute refers to one of three categories: first-year students, total students, or graduates (students who completed their degree in that year). The "grade" attribute denotes the educational level, which may be bachelor's, master's, or PhD.

Countries were first analyzed by clustering the countries based on their economic zone (results are presented in Section 3.1), followed by clustering according to their geographic area (results in Section 3.2) and their GDP, with results detailed in Section 3.3.

## 3.1 Analysis Made on the Economic Zones

Countries were first clustered based on their **Economic Zone.**

Figure 1 shows the female percentages of students enrolled in bachelor, master, and PhD programs across the three different European Economic Zones according to the classification used by

Informatics Europe, where Zone 0 identifies countries where the national GDP per capita (PPP) is greater than $65,000 (based on the World Bank's GDP per capita data for 2022). Zone 1 is for countries with a GDP per capita between $50,000 and $65,000, Zone 2 is for countries with a GDP per capita between $33,000 and $50,000, and Zone 3 is for countries with a GDP per capita below $33,000.

For bachelor programs, as depicted in Figure 1, the mean trends are similar across economic zones, although Economic Zone 0 had a lower mean female participation percentage in 2010, hovering around 13%, and has now a slightly higher (compared to other economic zones) percentage at around 20%. Although this trend is rising, all economic zones hover between 15% and 20% in female participation in bachelor's degrees in Informatics. Significant outliers are Belgium in Economic Zone 1, whose trend is significantly lower than other countries in the same zone, and Romania in Economic Zone 2, the only country to reach 30% female participation in bachelor degrees across all involved countries.

For master degrees, the trend is similar between Economic Zones 0 and 2, which also exhibit higher percentages for their bachelor counterparts, hovering between 20% and 25%. Interestingly, the same does not apply for Economic Zone 1, where the mean participation in master's degrees is very similar to the mean percentage for bachelor's degrees. Unfortunately, we don't have a complete– time series for Romania's master's degree, but Estonia stands out as a positive outlier, reaching more than 40% female participation in master's degrees. Lastly, for PhD programs, the number of complete time series is significantly lower, leading to more drastic variations.

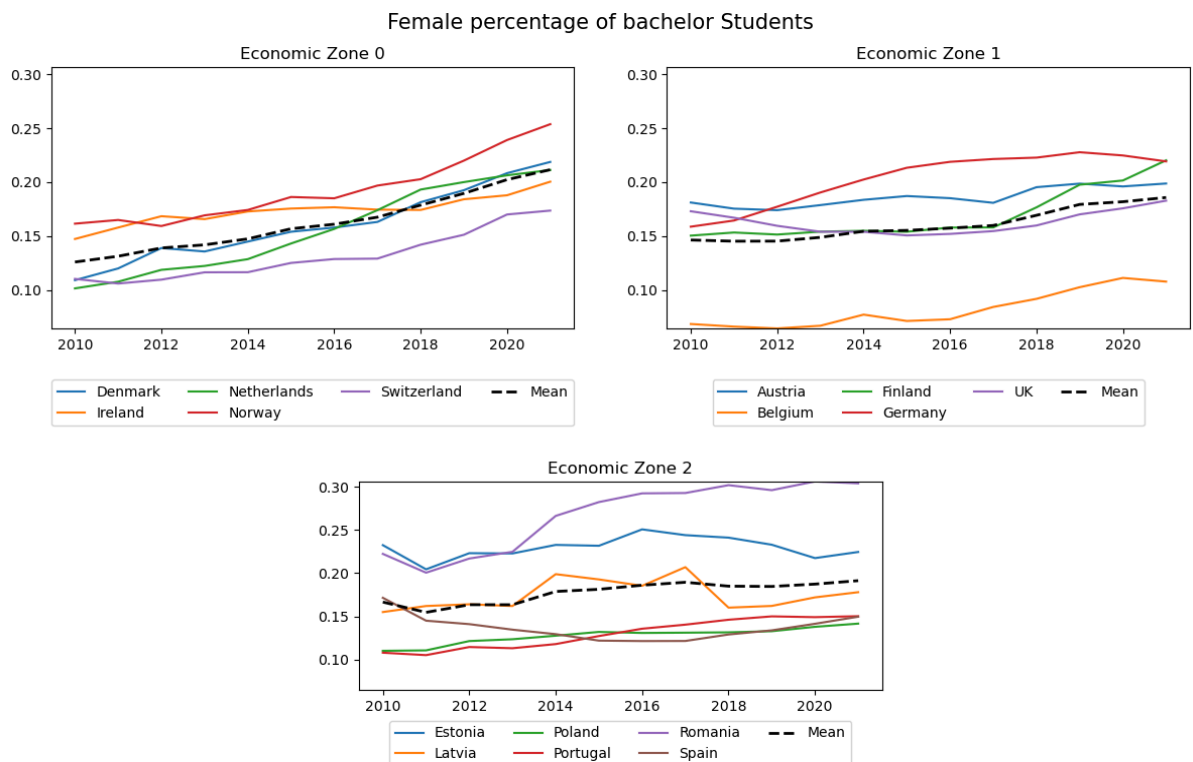

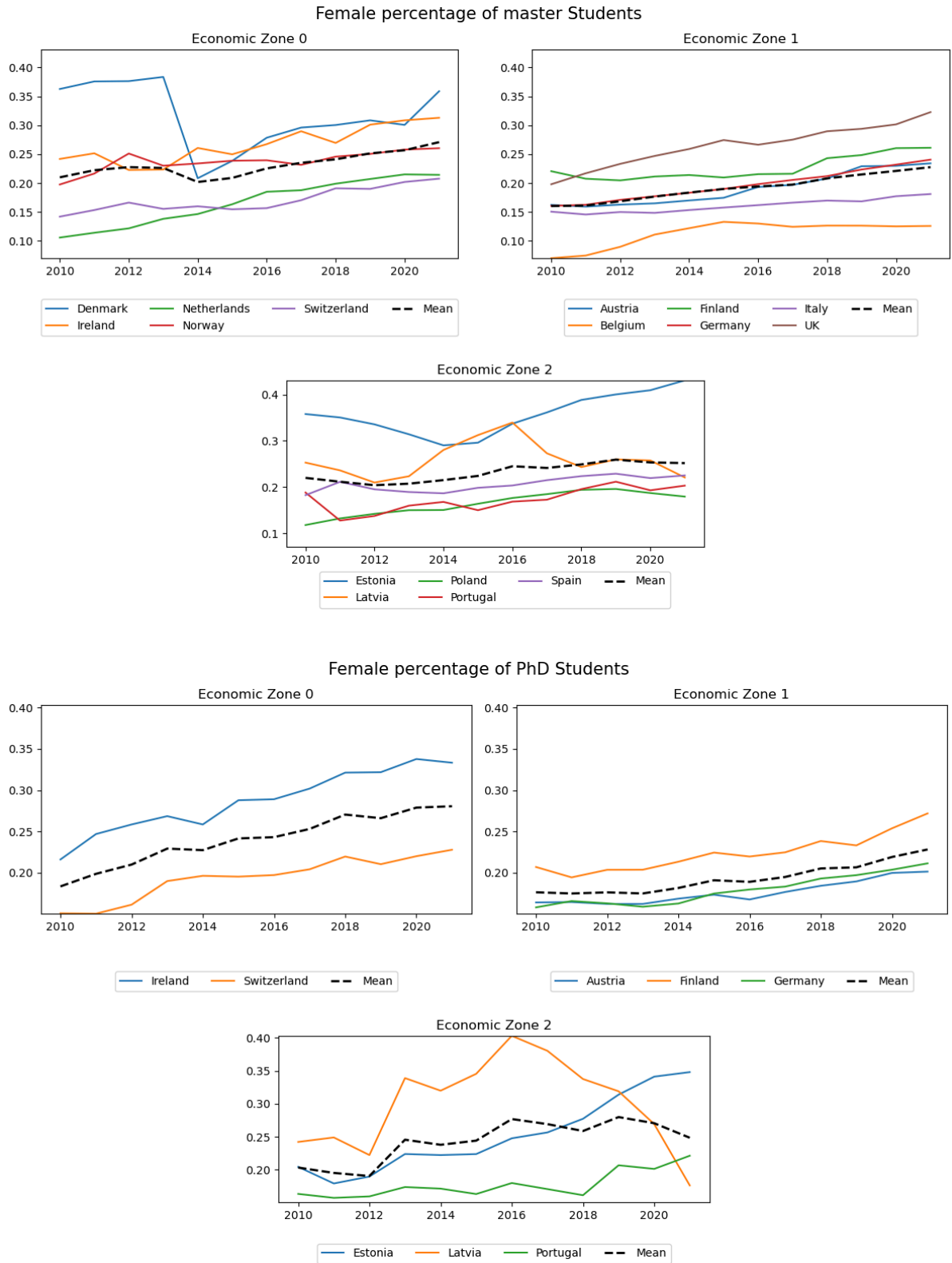

**Figure 1:** Female percentage of students enrolled in a. Bachelor, b. Master, and c. PhD Programs across the three Economic Zones.

For First Year students (Figure 2), the positive trend in Economic Zone 0 is noticeable. All economic zones are now set on a mean female percentage of first year students of around 20%. However, Economic Zone 2 only has two countries with complete data, so that it might be biased due to the low number of available samples.

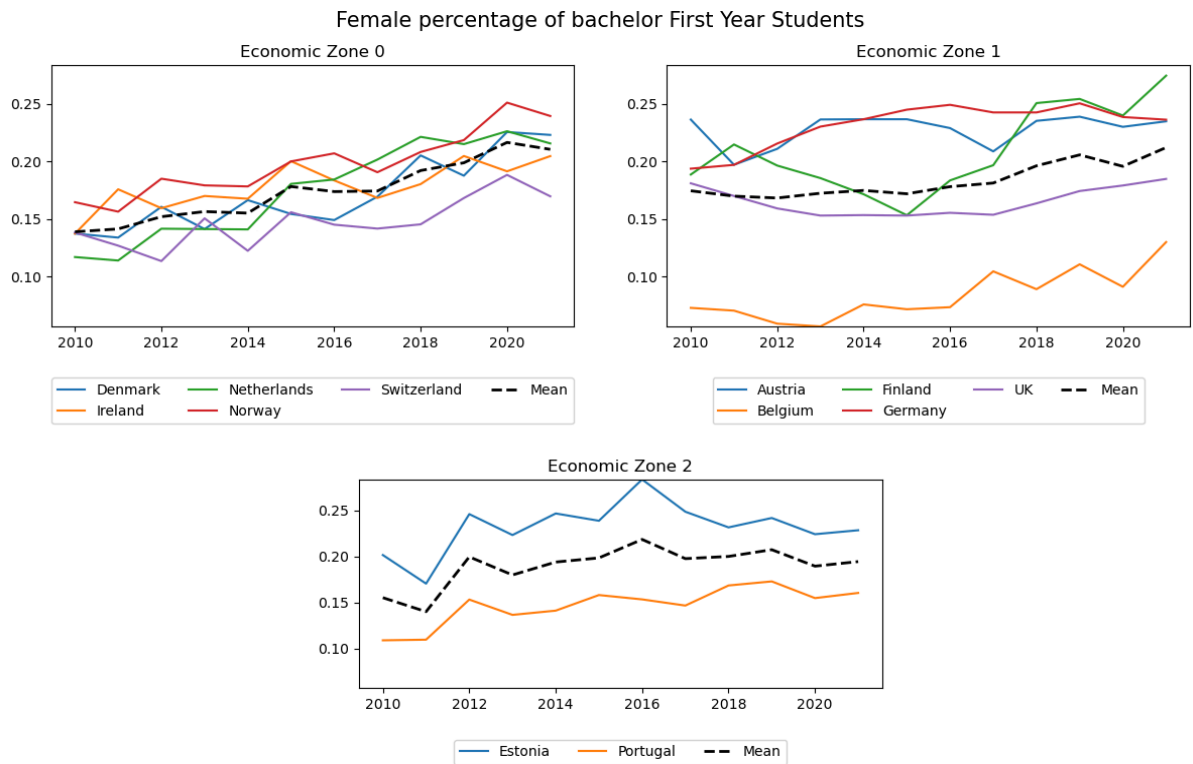

**Figure 2:** Female percentage of first year students enrolled in bachelor programs across the three Economic Zones.

Figure 3 shows how the number of Awarded bachelor degrees tells a similar story to the number of bachelor students in all Economic Zones. Economic Zones 0 and 1 are very similar, hovering between 15% and 20%, while Economic Zone 2 presents slightly higher percentages, between 20% and 25%, from 2010 to 2021.

For master's degrees, the temporal series are closer to each other, and all follow a generally positive trend: the mean percentage of master's degrees awarded to women is on the rise. The percentages are also generally much higher than those of their bachelor counterparts.

For the awarded PhDs, the available time series for Economic Area 0 are only two, so the results might be biased. Economic Zones 1 and 2 have similar trends, but the latter has higher percentages, with the mean hovering between 20% and 30%.

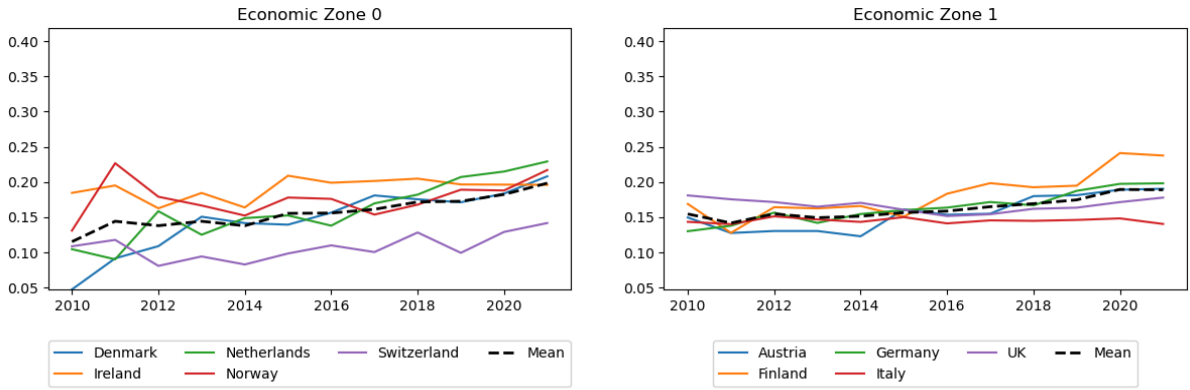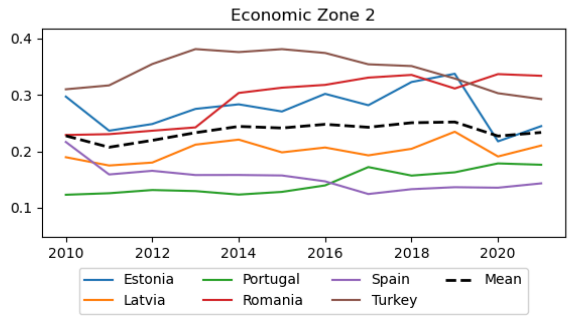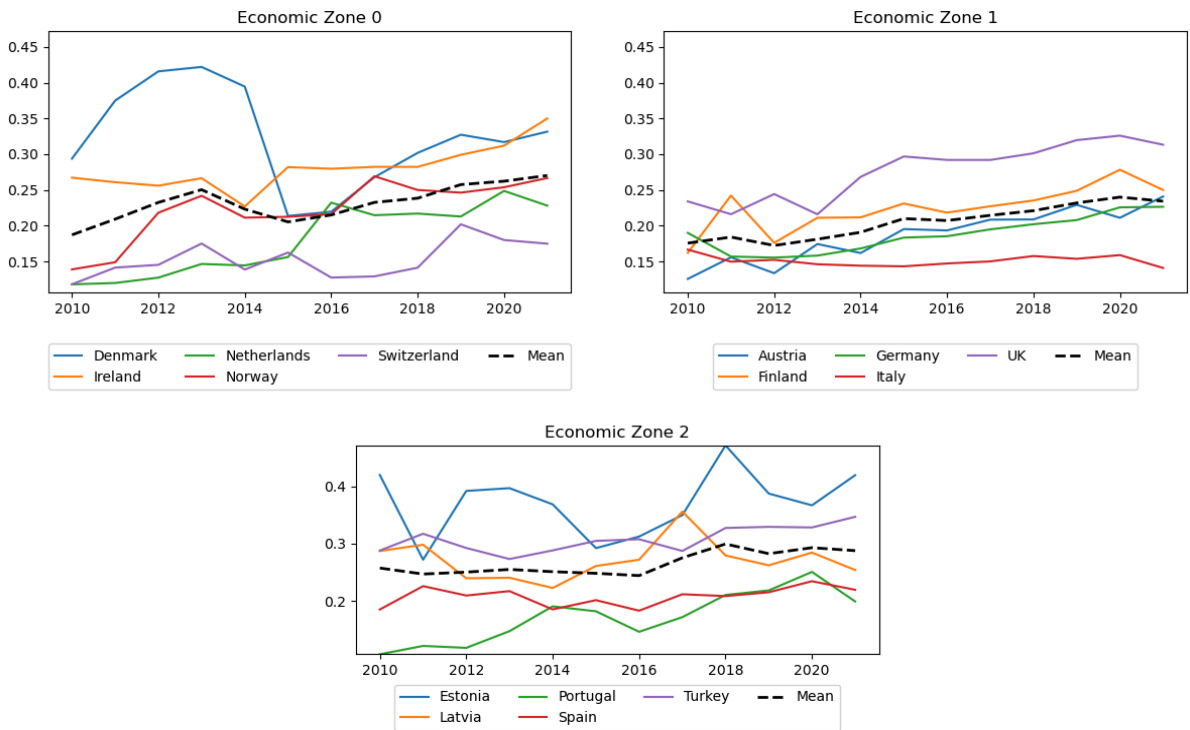

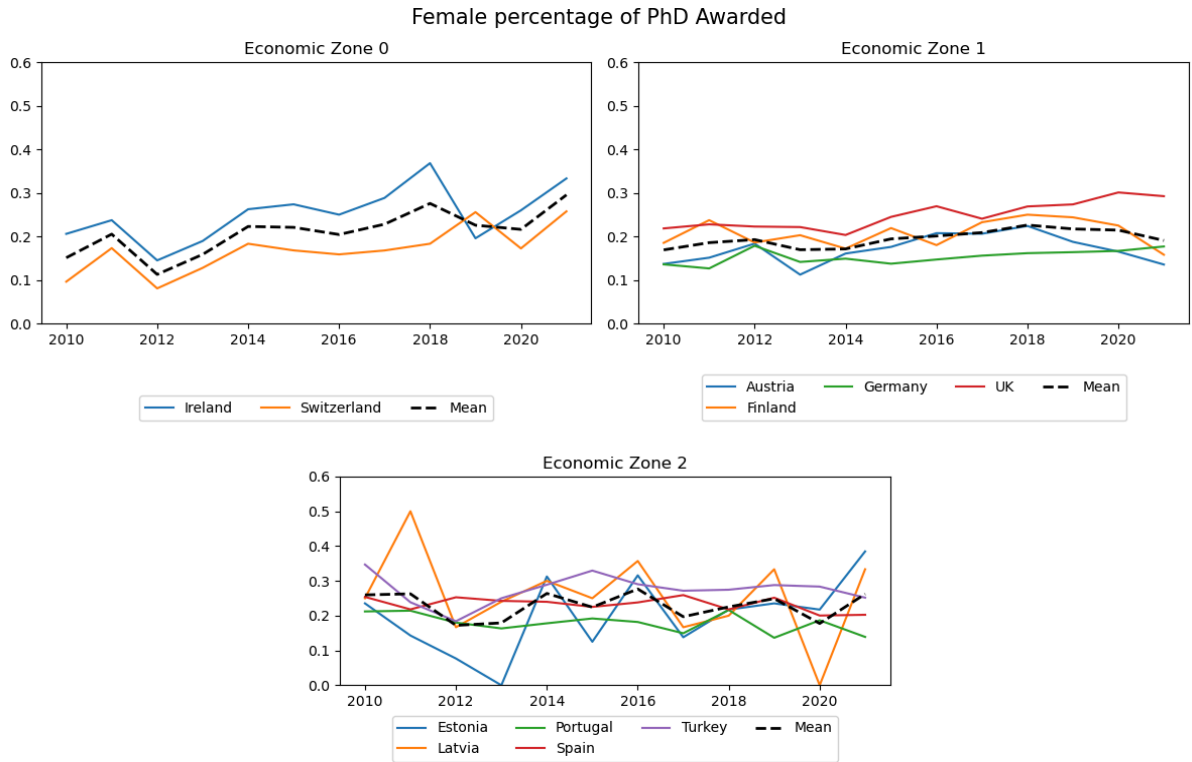

**Figure 3:** Female percentage of awarded degrees in a. Bachelor, b. Master, and c. PhD Programs across the three Economic Zones.

In conclusion, we cannot draw meaningful correlations between the economic zones and the female percentage of students or degrees awarded. However, Economic Zone 2 stands out generally with slightly higher (of around 5%) percentages, including big outliers like Romania and Latvia. This suggests that there might be a deeper correlation between GDP and female participation. We explore this clustering variable in Section 3.3.

Generally speaking, female participation in master's degrees is much higher than in bachelor's degrees. This observation is consistent across all Economic Zones.

## 3.2 Analysis Made in the Geographic Areas

The second clustering variable we employ to classify European countries is their Geographic Area. Each country is clustered in North, South, West, or East Europe (see Table 1 for the data sources).

The **Eastern zone** includes Bosnia and Herzegovina, Bulgaria, Cyprus, Czechia, Estonia, Georgia, Hungary, Israel, Lithuania, Moldova, Poland, Romania, Slovakia, and Ukraine.

The **Northern zone** comprises Denmark, Estonia, Finland, Iceland, Ireland, Latvia, Norway, Sweden, and the UK.

Within the **Southern zone**, we have Albania, Andorra, Armenia, Azerbaijan, Greece, Italy, Macedonia, Malta, Montenegro, Portugal, Serbia, Slovenia, Spain, and Turkey.

The **Western zone** comprises Austria, Belgium, France, Germany, Luxembourg, the Netherlands, and Switzerland.

Figure 4 shows the trends of the percentage of female students in bachelor, master, and PhD programs divided by the Geographic Areas of European countries. We have very few complete time series for the Eastern Geographic Area, which will be excluded from this commentary analysis.
For bachelor degrees, the mean percentage of female students in North Europe is consistently higher than in the other Areas from 2010 to 2021. This observation is exacerbated for master's degrees, where the percentages are generally higher across all areas. This is consistent with the results observed in the previous clustering of Economic Zones, discussed in Section 2.1.

The data available for PhD programs is scarce and not worth discussing. For North Europe, the mean trend is generally on the rise. Despite a small hiccup from 2020 to 2021, the percentage of female PhD students rose from around 22% to almost 30%.

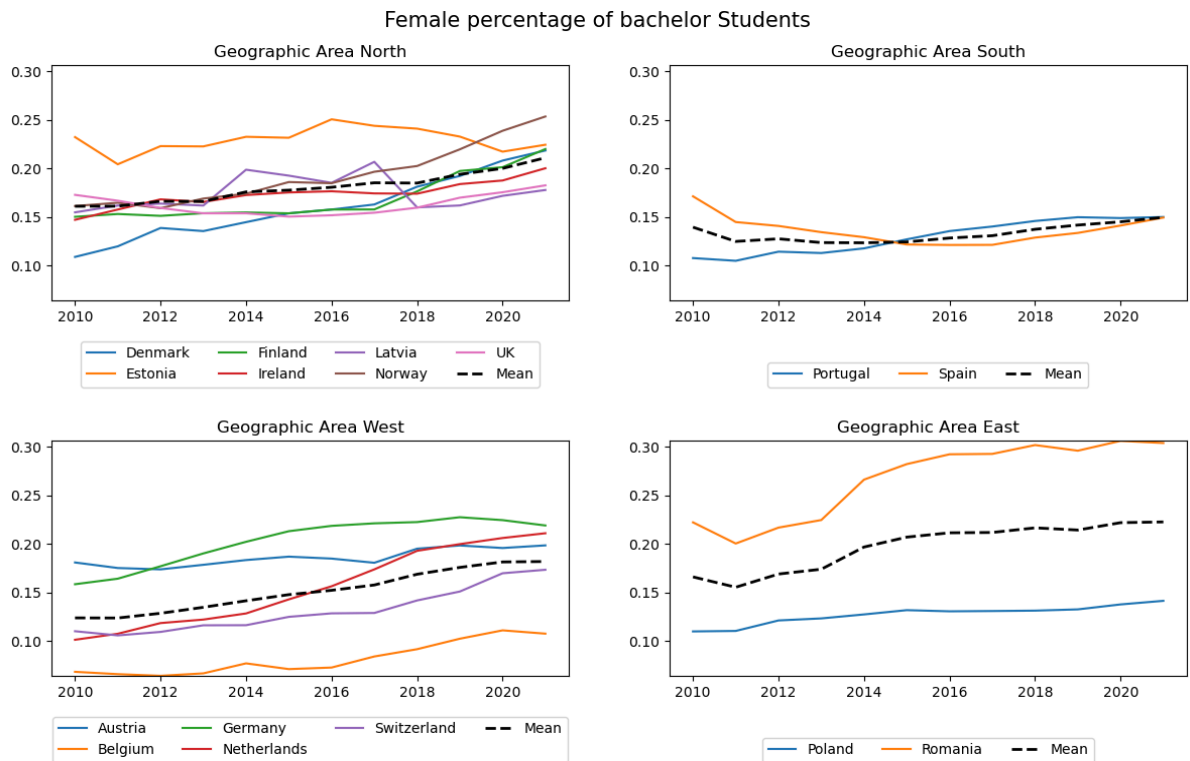

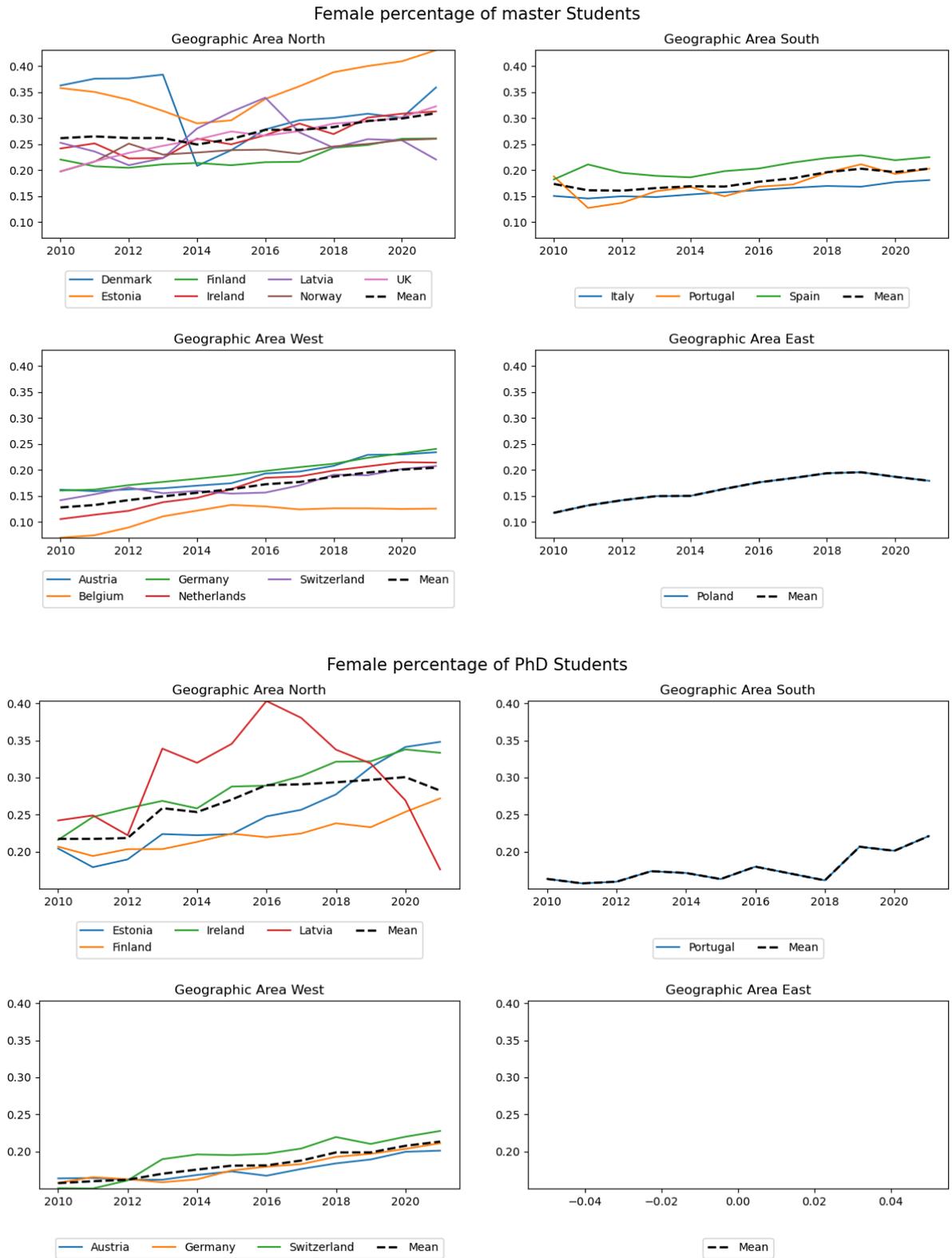

**Figure 4:** Female percentage of students enrolled in a. Bachelor, b. Master, and c. PhD Programs across the four Geographic Areas.

Figure 5 illustrates the percentage of female first-year students in bachelor's programs across different regions. Northern Europe exhibits a higher mean percentage than Western Europe, aligning with the trends observed in Figure 4, which shows a similar pattern for overall female participation in bachelor's degrees. However, due to the limited availability of Southern and Eastern Europe data, we cannot directly compare these regions, leaving potential regional variations in female participation unexplored.

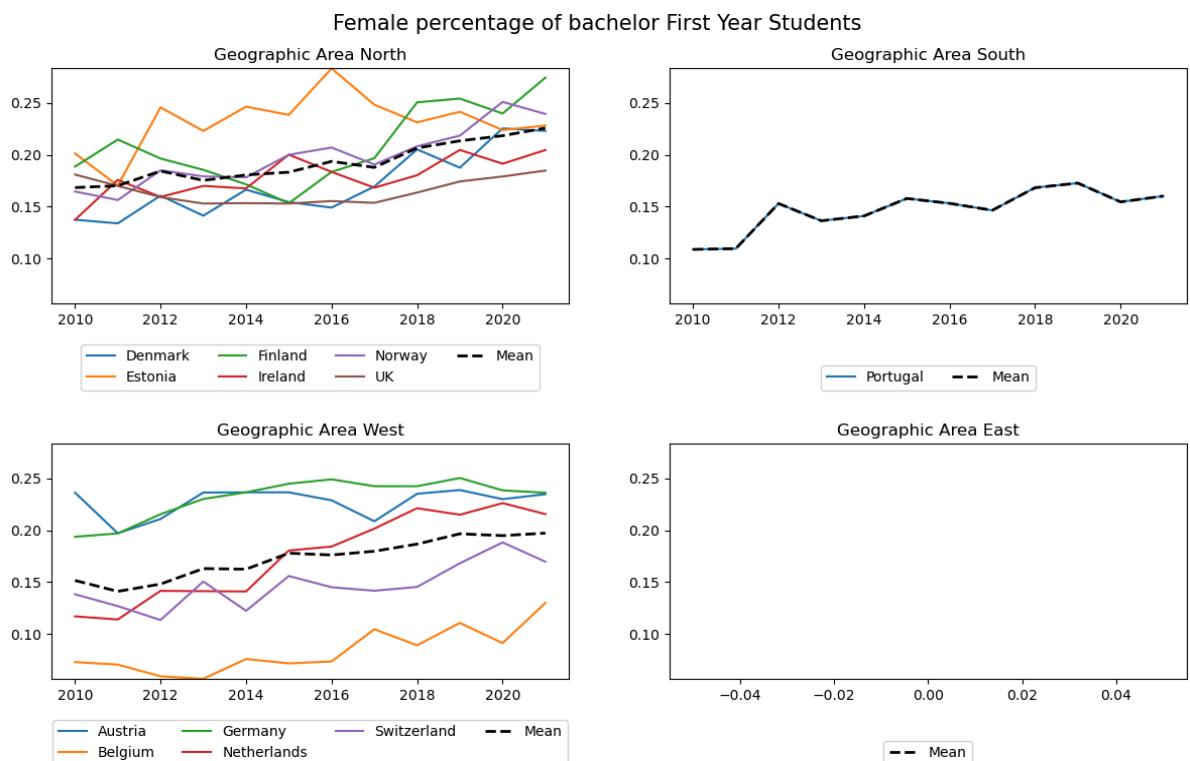

**Figure 5:** Female percentage of first year students enrolled in bachelor programs across the four Geographic Areas.

Lastly, Figure 6 shows the percentage of degrees awarded to women for each Geographic Area. For awarded bachelor degrees, North and South Europe display similar mean trends, though the latter is heavily influenced by Turkey, which stands out with significantly higher percentages. Both regions show stagnation, with little to no improvement in female participation over the past decade. In contrast, Western Europe shows notably lower percentages, while Eastern Europe has significantly higher values. However, the data for Eastern Europe is limited, as only two complete time series are available, making broader conclusions difficult.

For master's degrees, the mean trend for North Europe is higher than that of South and West Europe. There are no evident outliers in this case, so it is reasonable to conclude that Northern European countries enroll, in percentage, more female students in Informatics than other European geographical areas.

Lastly, for PhD programs, the trends of the individual countries appear to be less consistent and more erratic every year, especially in Northern countries. The mean trend, however, is extremely similar across all Geographic Areas, hovering between 20% and 30%.

## Female percentage of bachelor Awarded

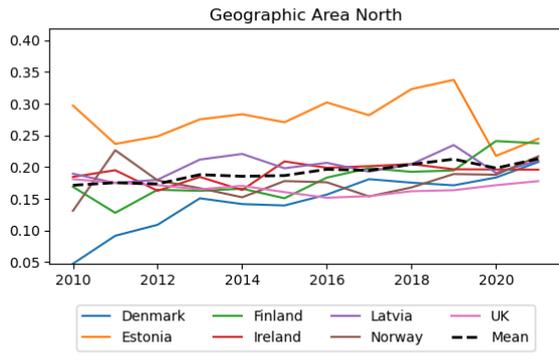
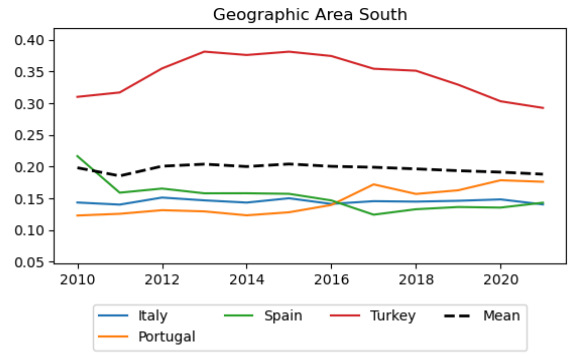
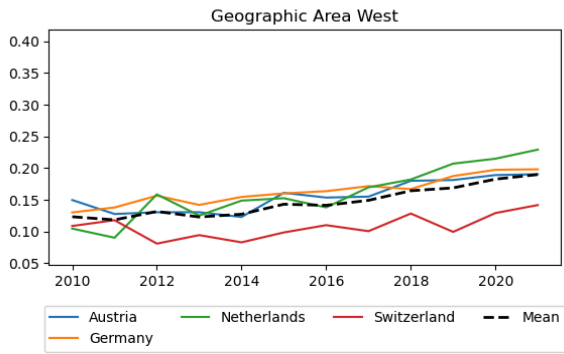
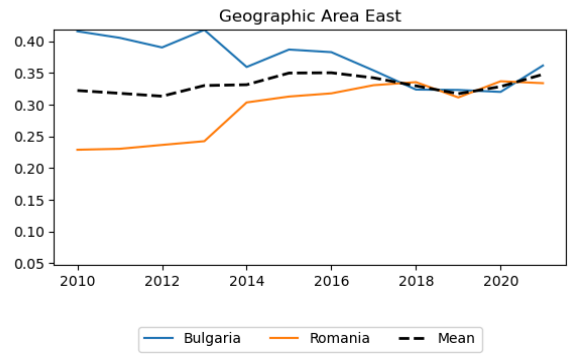

## Female percentage of master Awarded

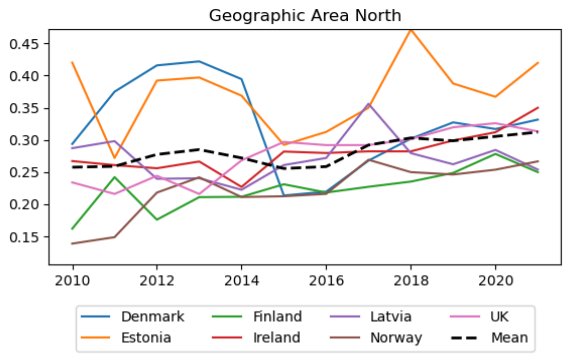
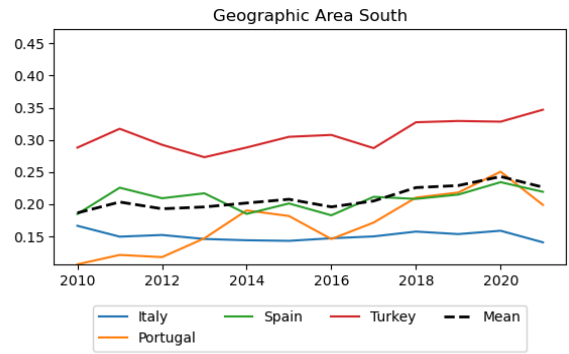
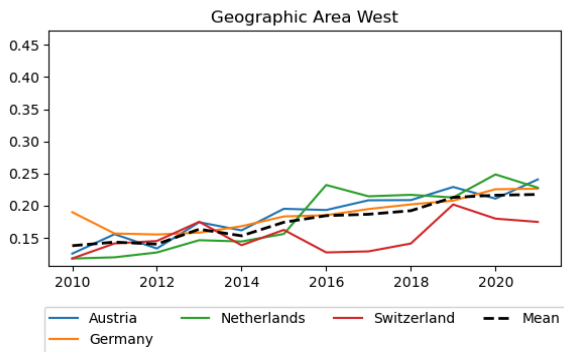
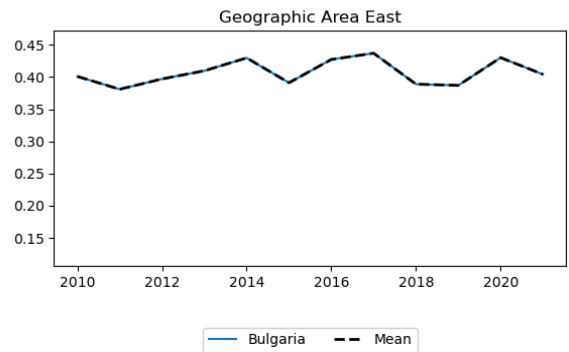

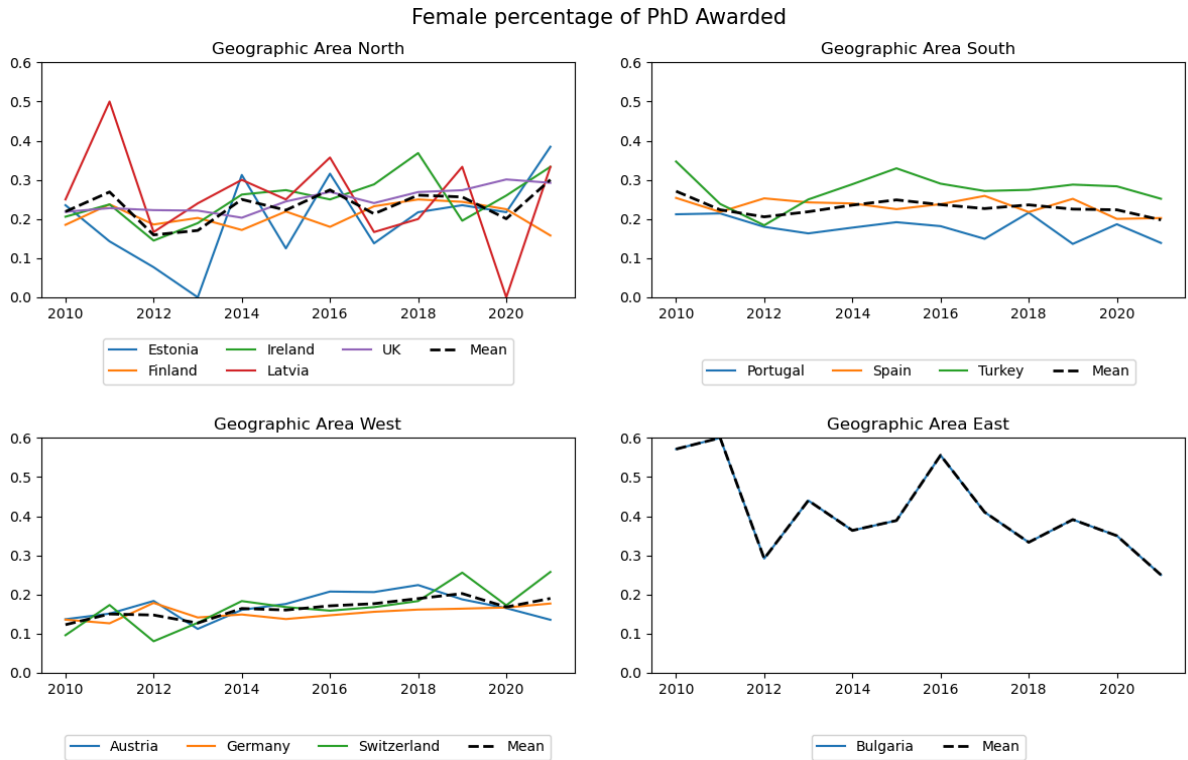

**Figure 6:** Female percentage of awarded degrees in a. Bachelor, b. Master, and c. PhD Programs across the four Geographic Areas.

## 3.3 Analysis Made on the GDP

Given that the results of the classification by economic zone suggested a potential relationship between female participation in Informatics and national GDP levels, we sought to further investigate this correlation by clustering European countries according to their GDP. Specifically, we classified the countries into quartiles, with Quartile 1 representing the lowest GDP and Quartile 4 representing the highest. Table 3 shows the minimum and maximum GDP for each quartile.

| Quartile | Min GDP | Max GDP | Countries |
| --- | --- | --- | --- |
| Q1 | 38689 | 48992 | Bulgaria, Estonia, Greece, Latvia, Portugal, Romania, Turkey |
| Q2 | 48992 | 58906 | Czechia, Italy, Lithuania, Poland, Spain, UK |
| Q3 | 58906 | 74485 | Austria, Belgium, Finland, France, Germany |
| Q4 | 74485 | 127623 | Denmark, Ireland, Netherlands, Norway, Switzerland |

**Table 3:** Min and Max GDP for each quartile.

This stratification allowed for a more detailed analysis of how female participation rates may vary across countries with differing economic capacities, thereby providing deeper insights into the potential influence of economic factors on gender representation in Informatics.

Figure 7 presents the percentage of female students enrolled in Bachelor's, Master's, and PhD programs across countries categorized by GDP quartiles. For Bachelor's degrees, countries in the lowest GDP quartile (Quartile 1) generally exhibit higher female participation rates than other quartiles, which display more consistent trends. Notably, Quartile 1 also demonstrates stronger female representation in Master's programs despite the overall mean being somewhat lowered by the influence of Portugal.

For PhD programs, we do not have any complete data for countries in Quartile 2. Countries in the other quartiles show similar results, but notably, Latvia has fallen from one of the best-performing countries to one of the worst during the last five years, influencing the mean trend of countries in Quartile 1.

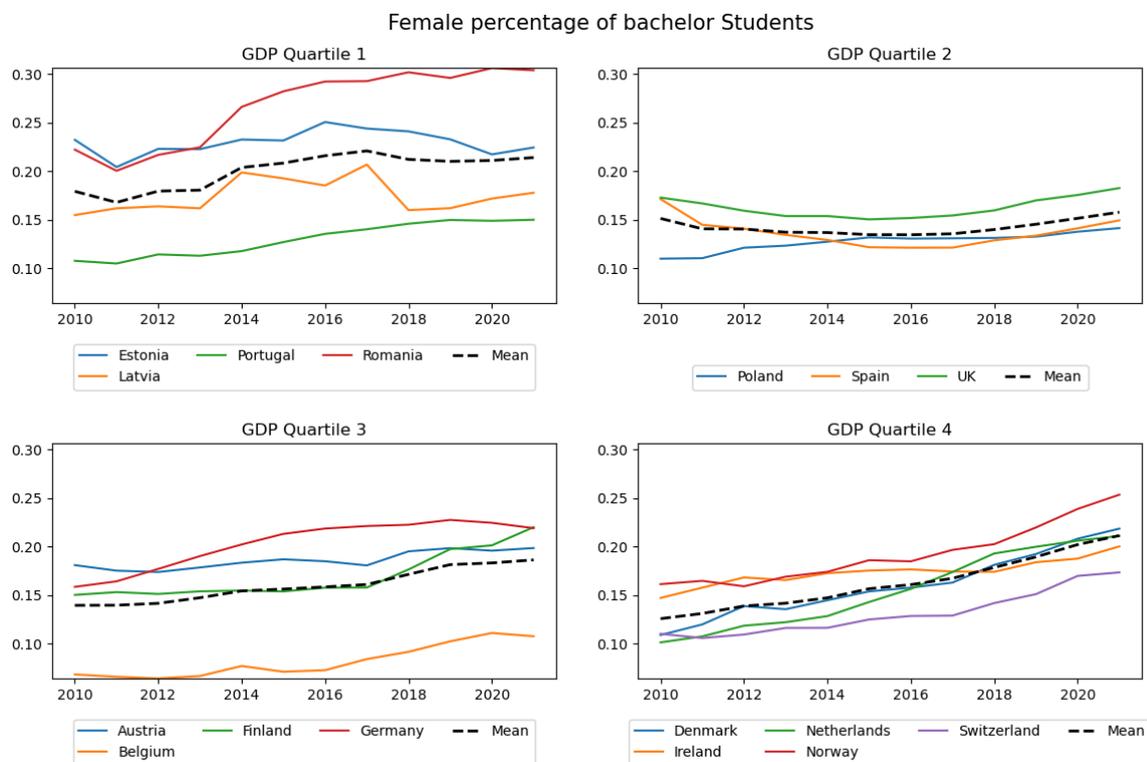

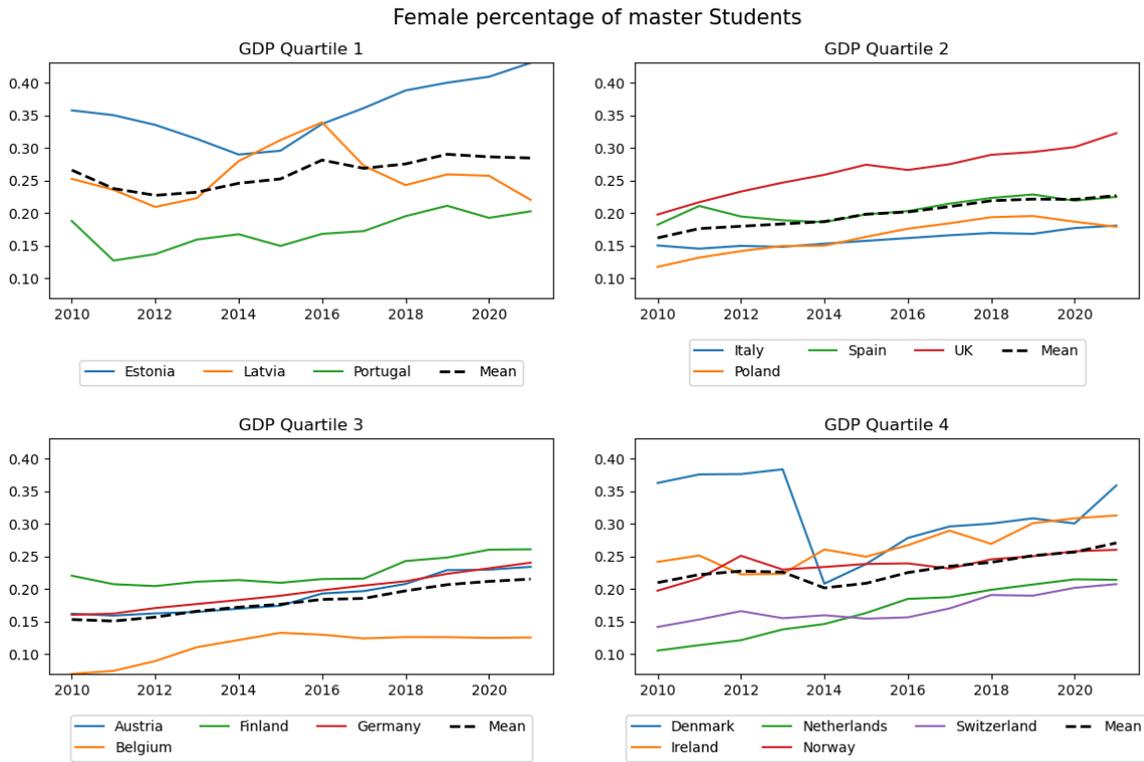

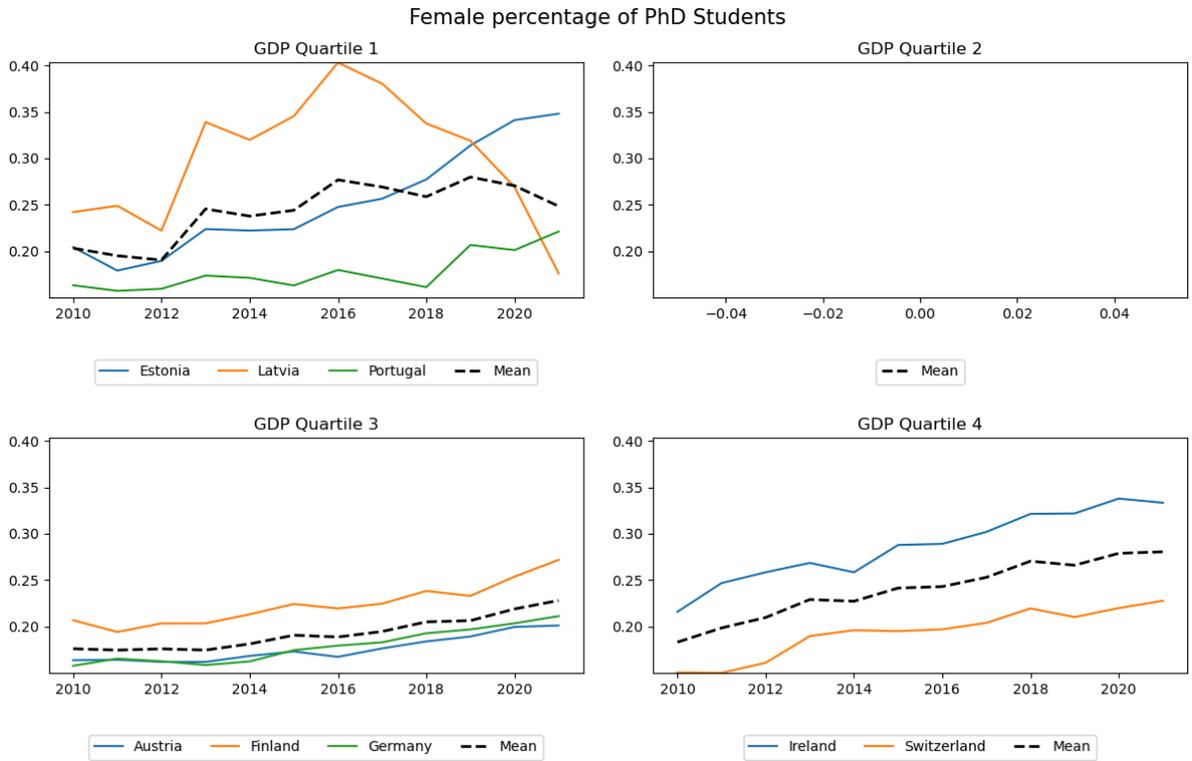

**Figure 7:** Female percentage of students enrolled in a. Bachelor, b. Master, and c. PhD Programs across the four GDP Quartiles.

Continuous data for female percentages of newly enrolled students, shown in Figure 8, is especially lacking for Quartiles 1 and 2 countries, making it difficult to draw meaningful conclusions from them. The trends are generally improving, especially for countries with the highest GDP (Quartile 4), which expose a consistent upward trend over the last decade.

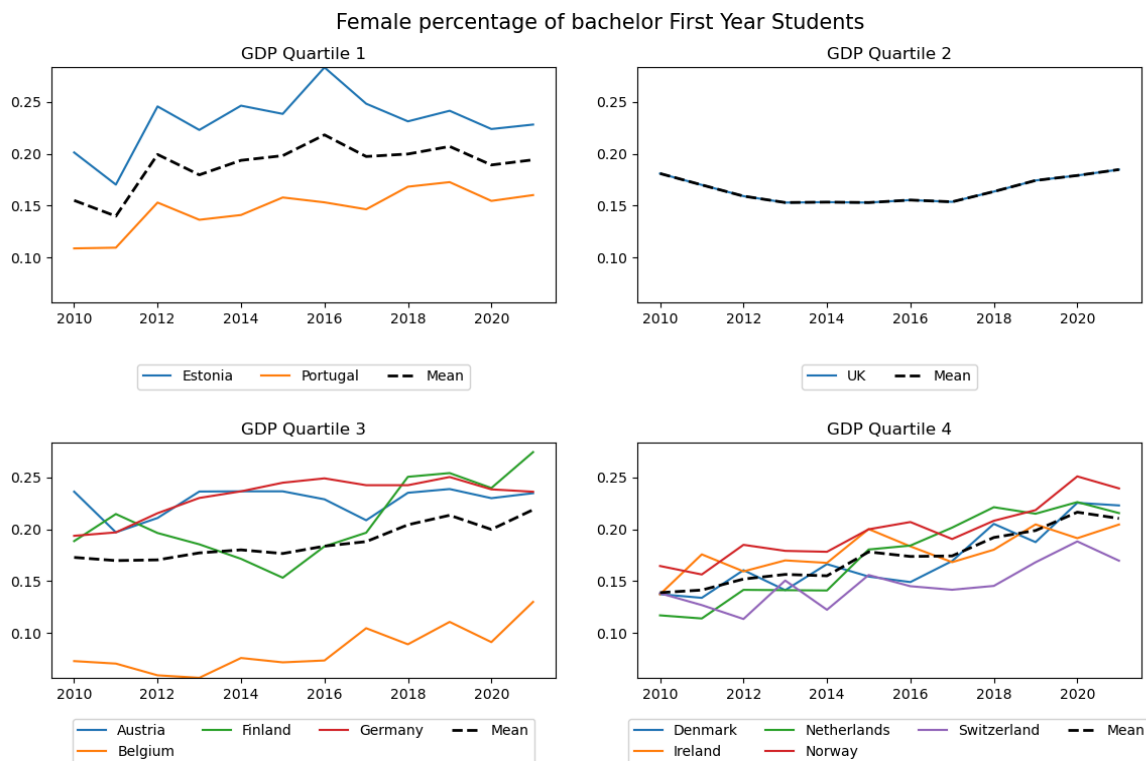

**Figure 8:** Female percentage of first year students enrolled in bachelor programs across the four GDP Quartiles.

Lastly, Figure 9 highlights the percentage of degrees awarded to female students. The stark gap between the mean for countries in the lowest GDP quartile (Quartile 1) and those in higher quartiles is particularly striking, despite Portugal acting as a negative outlier. These results demonstrate that countries with the lowest GDP consistently have a significantly higher percentage of female students receiving degrees, regardless of level (Bachelor's, Master's, or PhD). Across all quartiles, the mean trends appear either stagnant or showing slight improvement, with no evidence of decline, except for PhD awards in countries within the third quartile, where a low sample size may have introduced bias.

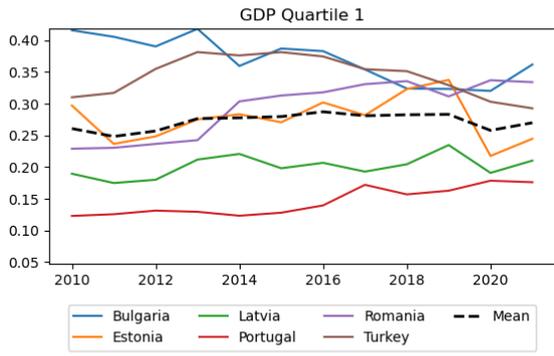
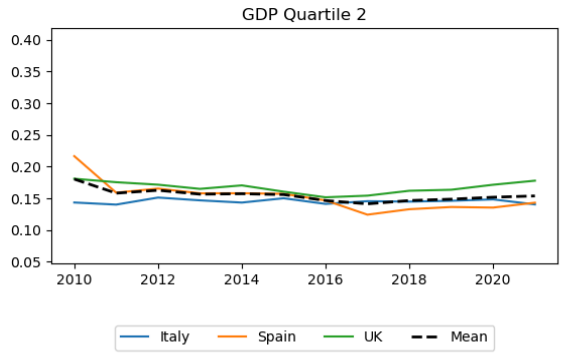
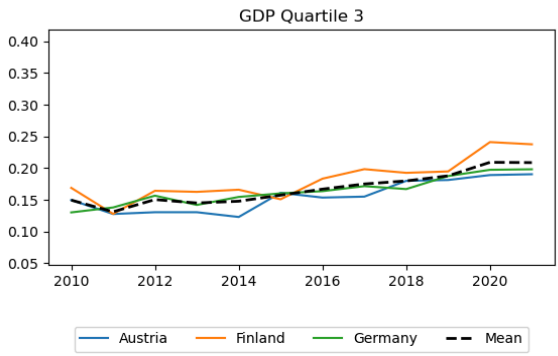
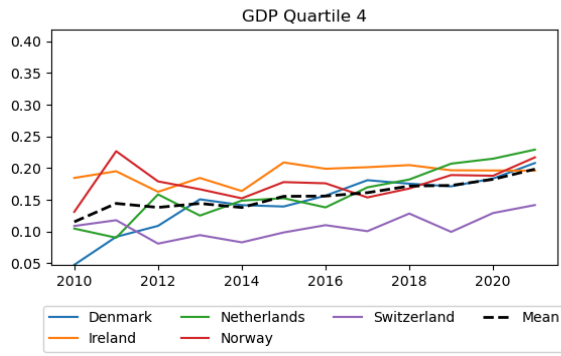
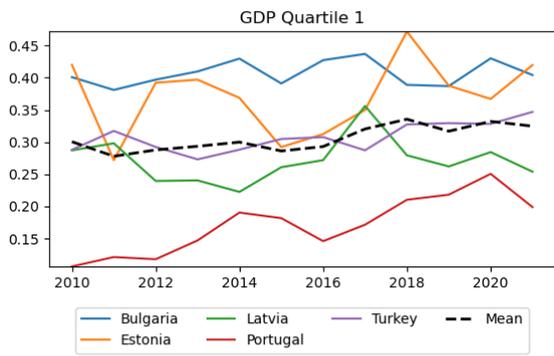
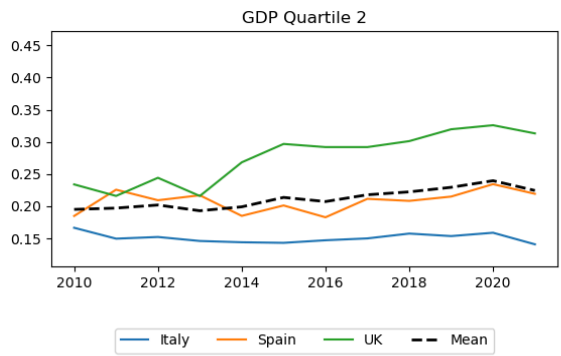
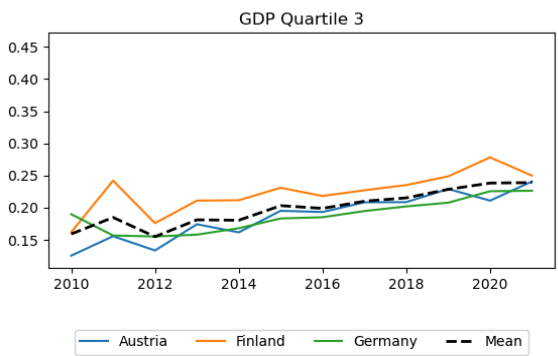
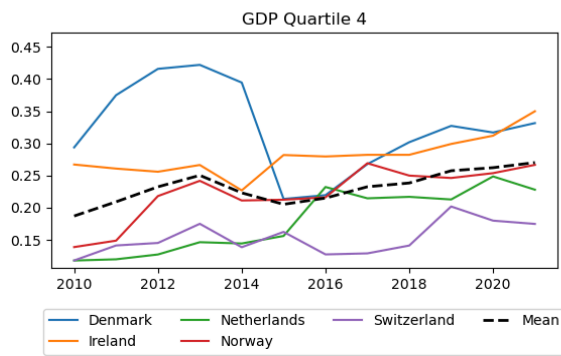

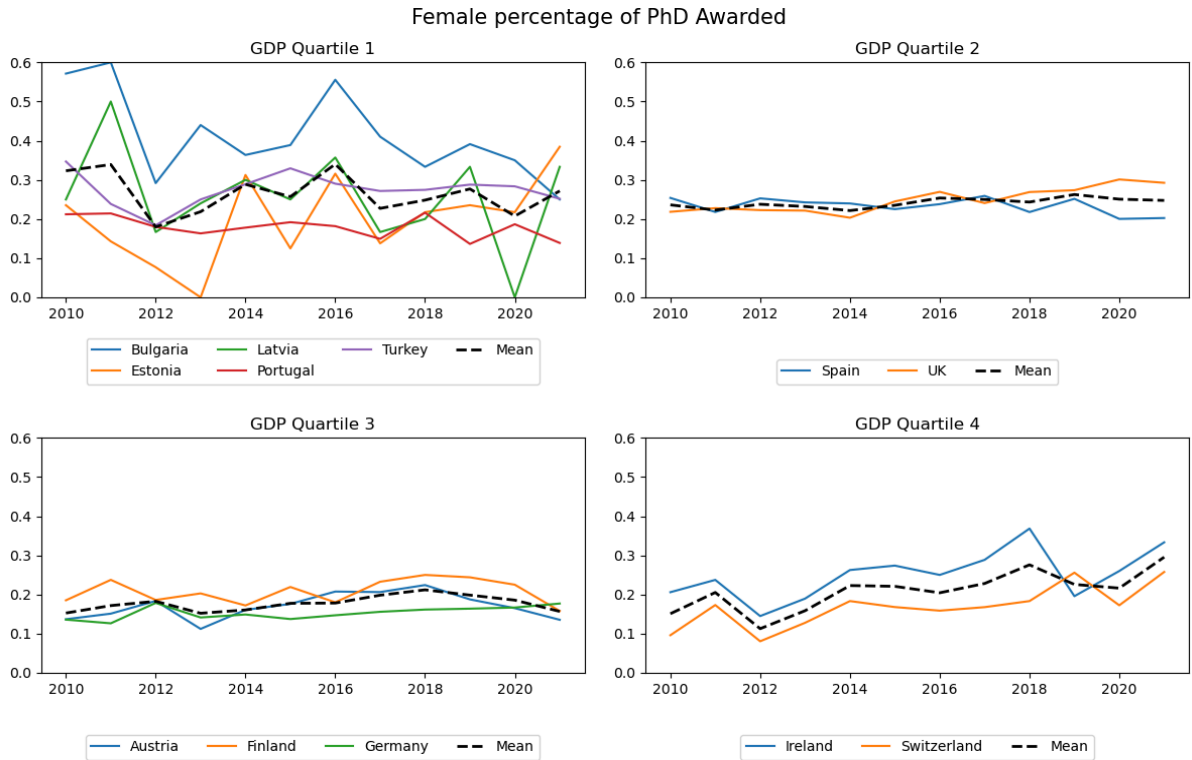

**Figure 9:** Female percentage of awarded degrees in a. Bachelor, b. Master, and c. PhD Programs across the four GDP Quartiles.

## 4. Conclusion

In this Virtual Mobility Grant, we enriched the Informatics Europe's Higher Education (IEHE) data portal dataset by incorporating several new features from various online sources. After cleaning the data to remove incomplete time series excluded from further analysis, we examined temporal trends by classifying countries using three distinct clustering variables: Economic Zone (as defined by the Informatics Europe website), Geographic Area, and GDP Quartiles. All of these clusterings were only possible through the augmentation of the dataset via collecting the data available online. While meaningful conclusions could not be drawn from comparisons based on Economic Zone or Geographic Area, examining outliers within these groups suggested that analysis of GDP Quartile might provide more insightful results. This was the case, as we identified a striking and consistent pattern: countries in the lowest GDP quartile consistently exhibited the highest percentage of female students and degrees awarded to women across all academic levels (Bachelor's, Master's, PhD). Though counterintuitive, these findings align with previous literature [1]. Breda et al. (2020) found that gender stereotypes, such as the belief that "math is not for girls," are stronger in wealthier, gender-equal countries, which could explain why lower-GDP countries show higher female participation in STEM, as these stereotypes are less pervasive [5]. Lastly, Falk and Hermle (2018) demonstrated that as wealth and gender equality increase, gender-differentiated preferences also increase. Economic pressures might drive women toward STEM fields in poorer countries, while in richer countries, women may choose other fields more freely [6].

In addition to these results, the project produced an enriched dataset, which can serve as a valuable resource for future research.